\documentclass[12pt]{article}
\usepackage{amsfonts}
\usepackage{amsmath}
\usepackage{amssymb}
\usepackage{graphicx}
\usepackage{color}
\usepackage[left=1cm,top=1cm,bottom=1cm,right=1cm,nohead,nofoot]{geometry}

\def \be {\begin{equation}}
\def \ee {\end{equation}}
\def \bea {\begin{eqnarray}}
\def \eea {\end{eqnarray}}
\def \nn {\nonumber}

\def \rr {\raise.35ex\hbox{\small $\prime$}\kern-.17em{\mbox{\large $\imath$}}}
\def \del {\partial}
\def \dels {\partial\kern-.6em /\kern.1em}
\def \As {{A\kern-.5em / \kern.5em}}
\def \Ds {D\kern-.7em / \kern.5em}
\def \a {\alpha}

\def \b {\beta}

\def \G {\Gamma}
\def \d {\delta}
\def \eps {\epsilon}

\def \ks {k\kern-.5em /}
\def \ls {l\kern-.5em /}\def \lam {\lambda}
\def \Lam {\Lambda}

\def \II {I\hspace{-.1em}I\hspace{.1em}}

\def \IIA {\mbox{\II A\hspace{.2em}}}

\def \dm {\dot{\mu}}
\def \dn {\dot{\nu}}

\def \dlam {\dot{\lambda}}

\reversemarginpar
\newcommand{\ol}{\overline}

\begin{document}
\begin{titlepage}

\begin{center}

\hfill
\vskip .2in

\textbf{\LARGE
BPS States on M5-brane in \\
\vskip.5cm
Large $C$-field Background
}

\vskip .5in
{\large
Pei-Ming Ho$\,{}^{a,b,c,}$\footnote{e-mail address: pmho@phys.ntu.edu.tw}
Chen-Te Ma$^{a,}$\footnote{e-mail address: yefgst@gmail.com} and\
Chi-Hsien Yeh$^{a,}$\footnote{e-mail address: d95222008@ntu.edu.tw}}\\
\vskip 3mm
{\sl
${}^a$
Department of Physics and Center for Theoretical Sciences, \\
${}^b$
Center for Advanced Study in Theoretical Sciences, \\
${}^c$
National Center for Theoretical Sciences, \\
National Taiwan University, Taipei 10617, Taiwan,
R.O.C.}\\
\vskip 3mm
\vspace{60pt}
\end{center}
\begin{abstract}

We extensively study BPS solutions of the low energy effective theory
of M5-brane in large $C$-field background.
This provides us an opportunity to explore
the interactions turned on by $C$-field background 
through the Nambu-Poisson structure.
The BPS states considered in this paper include
the M-waves,
the self-dual string (M2 ending on M5),
tilted M5-brane,
holomorphic embedding of M5-brane
and the intersection of two M5-branes along a 3-brane.

\end{abstract}

\end{titlepage}

\section{Introduction}
\label{1}

As M theory unifies all superstring theories \cite{Witten:1995ex},
various D-branes in superstring theories correspond to
M2-branes and M5-branes in M theory \cite{Townsend:1995kk}.
Analogous to how D-branes play a crucial role in string theories,
M2-branes and M5-branes are also the key ingredients in M theory,
and have been a focus of research interest.

A recent progress made in this direction was
the construction of the low energy effective theory
for a single M5-brane in the large $C$-field background \cite{M51,M52}.
\footnote{
The large $C$-field limit for M5-brane was first considered in 
\cite{Bergshoeff}.
}
A salient feature of this model is the Nambu-Poisson structure
which dictates the gauge symmetry and interactions.
We will review this theory in the next section,
and refer to this theory as the Nambu-Poisson M5-brane theory,
or NP M5-brane theory in short.
Although the ordinary M5-brane theory \cite{ordinary-M5} can also 
describe $C$-field background, 
the difference lies in the way $C$-field is scaled 
together with other parameters in the low energy limit
(see eqs.(\ref{limit_lp})--(\ref{limit_C}) below).
The relation between NP M5-brane theory and the ordinary M5-brane theory 
is analogous to the relation between the noncommutative D-brane action
and the DBI-action \cite{Seiberg:1999vs}

The purpose of this work is to give an extensive
(but not exhaustive) search of BPS states 
that preserve one half of the supersymmetry
in the NP M5-brane theory, 
as a further step to understand the physics of M5-brane 
in large $C$-field background.



The plan of this paper is as follows.
We review the NP M5-brane theory in Sec. \ref{2}.
In Sec. \ref{3} we systematically study BPS configurations.
We organize the BPS solutions according to the number of 
scalar fields that are turned on, that is,
the number of transverse directions on the M5-brane 
that have nontrivial fluctuations.
The light-like BPS states that are pure gauge field configurations
are given in Sec. 3.1.
There we observe the interesting fact
that there exist static configurations
that do not satisfy equations of motion,
but they satisfy all BPS conditions 
(i.e. they preserve partial supersymmetry).
We will briefly comment on the validity of the folklore that 
BPS conditions imply equations of motion.
In Sec. 3.2, we turn on a single scalar field 
and describe self-dual string solutions which 
represent M2-branes ending on M5-branes along 
a one-dimensional brane.
Due to the $C$-field background, 
rotation symmetry on the M5-brane is broken 
and the self-dual string solution depends on the direction 
along which it extends.
We also shortly comment on a solution representing 
a tilted M5-brane, 
on which the gauge field strength has to be turned on 
to preserve SUSY
due to the $C$-field background.
In Sec. 3.3, 
we study BPS states for which the M5-brane 
is a K\"{a}hler manifold in the spacetime.
This class of solutions include the special case 
representing the intersection of two M5-branes along 
a 3-dimensional brane.
We find that the gauge field has to be turned on 
if two or more of the transverse directions 
are excited on the M5-brane with nontrivial dependence 
on the 3 directions chosen by the $C$-field background.
Finally, in Sec. \ref{4} we conclude with remarks 
on potential future research directions.

\section{Review of M5-brane in $C$-field background}
\label{2}

Analogous to how a D$(p+2)$-brane with a background flux
can be constructed from infinitely many D$p$-branes
\cite{DpDp+2},
an M5-brane can be constructed from infinitely many M2-branes.
The low energy effective theory of a single M5-brane in
large $C$-field background \cite{M51,M52} can thus be
derived from the Bagger-Lambert action \cite{BL}
for multiple M2-branes in M theory
by choosing the Nambu-Poisson algebra as the Lie 3-algebra
of gauge symmetry.
We will refer to this theory as the ``NP M5-brane theory''
and give a brief review in this section.

The NP M5-brane theory is a good description of the M5-brane 
in $C$-field background in the simultaneous low energy and large $C$-field limit
defined by the scaling relations
\cite{Chen:2010br}
\bea
&\ell_P \sim \eps^{1/3},
\label{limit_lp} \\
&g_{(M)\mu\nu} \sim \eps^{0},
\qquad
g_{(M)\dot\mu\dot\nu} \sim \eps
\label{limit_g} \\
&C_{\dm\dn\dlam} \sim \eps^0
\label{limit_C}
\eea
with $\eps \rightarrow 0$,
where $\ell_P$ is the Planck length
and $g_{(M)}$ is the spacetime metric.
The convention of the indices here are that
the dotted indices ($\dm, \dn = \dot1, \dot2, \dot3$)
label the directions in which the $C$-field component dominates.
The remaining three worldvolume directions are
labeled by $\mu, \nu = 0, 1, 2$, which
are the worldvolume directions of the infinitely many M2-branes constituting the M5-brane.
The component $C_{\mu\nu\lam}$ (essentially just $C_{012}$)
is determined by $C_{\dot1\dot2\dot3}$ through the nonlinear self-duality condition
on the M5-brane.
The scaling limit eqs.(\ref{limit_lp})--(\ref{limit_C})
implies that $C_{012}$ is negligible 
in comparison with $C_{\dot1\dot2\dot3}$
\cite{Chen:2010br}.

This scaling limit defined by eqs.(\ref{limit_lp})--(\ref{limit_C})
is analogous to the zero-slope limit of Seiberg and Witten
\cite{Seiberg:1999vs} in which the low energy effective theory
of a D-brane in large $B$-field background can be conveniently
described as a noncommutative gauge theory.
Indeed the Nambu-Poisson gauge symmetry is analogous
to the noncommutative gauge symmetry,
but it is not a full-blown generalization of the latter.
Instead it is an extension of the Poisson structure,
which is the leading order approximation of noncommutative structure
in the $1/B$-expansion.

The low-energy effective theory of a D4-brane
in \IIA superstring theory 
should be related to that of the M5-brane
via double dimensional reduction (DDR). 
In Ref. \cite{M52}, it was shown that 
when one of the directions of $x^{\dm}$ is compactified, 
the NP M5-brane theory reduces to the low-energy effective theory
for D4-brane in the large $B$-field background.
More recently, the low energy effective theory for a D4-brane in large $C$-field background 
was derived from the NP M5-brane theory 
via DDR along the direction of $x^1$ or $x^2$ \cite{Ho}.
The BPS states studied in this paper should have their counterparts
on a D4-brane in large $C$-field background \cite{Ho:2012}.

\subsection{Action}

The M5-brane theory has the ${\cal N} = (0, 2)$ supersymmetry in 6 dimensions.
The field content of the low energy effective theory
is composed of a 2-form potential $b_{\tilde\mu\tilde\nu}$
($\tilde\mu, \tilde\nu = 0, 1, 2, \dot1, \dot2, \dot3$),
5 scalars $X^I$ ($I = 6, 7, 8, 9, 10$)
and their fermionic superpartner $\Psi$,
which is half of an 11-dimensional Majorana spinor,
equivalent to two 6-dimensional Weyl spinors.
An important feature of the theory is that the gauge field sector
is a self-dual gauge theory,
hence the number of independent polarizations of the 2-form potential is 3,
and the on-shell degrees of freedom for both bosons and fermions are
8 times that of a scalar.

In the following,
the signature of spacetime is given as $\eta=\mbox{diag}(-+\cdots +)$
in our convention.
The linearized action for an M5-brane in a large $C$-field background was
found in \cite{M51},
and then the complete nonlinear version in \cite{M52}.
It is
\be
S = \frac{T_{M5}}{g^2}
\left( S_{\mbox{\em \small boson}} + S_{\mbox{\em \small fermi}} + S_{CS} \right),
\label{M5S}
\ee
where $T_{M5}$ is the M5-brane tension and
\footnote{
$\Psi$ here was denoted by $\Psi'$ in \cite{M52}.
It is chiral, i.e., $\Gamma^7 \Psi = \Psi$,
where $\Gamma^7$ is chirality operator in 6 dimensions
defined by
$\Gamma^7 \equiv \Gamma^{012\dot{1}\dot{2}\dot{3}}$.
}
\begin{eqnarray}
S_{\mbox{\em \small boson}}
&=&\int d^6 x  \; \left[
-\frac{1}{2}({\cal D}_\mu X^I)^2
-\frac{1}{2}({\cal D}_{\dot\mu}X^I)^2
-\frac{1}{4}{\cal H}_{\mu\dot\mu\dot\nu}^2
-\frac{1}{12}{\cal H}_{\dot\mu\dot\nu\dot\lambda}^2
\right.\nonumber\\&&\left.
-\frac{1}{2g^2}
-\frac{g^4}{4}\{X^{\dot\mu},X^I,X^J\}^2
-\frac{g^4}{12}\{X^I,X^J,X^K\}^2\right],
\label{Sboson}
\\
S_{\mbox{\em \small fermi}}
&=&\int d^6 x  \; \left[
\frac{i}{2}\ol\Psi\Gamma^\mu {\cal D}_\mu\Psi
+\frac{i}{2}\ol\Psi\Gamma^{\dot\mu}{\cal D}_{\dot\mu}\Psi
\right.\nonumber\\&&\left.
+\frac{ig^2}{2}\ol\Psi\Gamma_{\dot\mu I}\{X^{\dot\mu},X^I,\Psi\}
-\frac{ig^2}{4}\ol\Psi\Gamma_{IJ}\Gamma_{\dot1\dot2\dot3}\{X^I,X^J,\Psi\}
\right],
\label{Sfermi0}
\\
S_{CS}
&=&
\int d^6 x  \;
\epsilon^{\mu\nu\lambda}\epsilon^{\dot\mu\dot\nu\dot\lambda}
\left[ -\frac{1}{2}
\partial_{\dot\mu}b_{\mu\dot\nu}\partial_\nu b_{\lambda\dot\lambda}
+\frac{g}{6}
\partial_{\dot\mu}b_{\nu\dot\nu}
\epsilon^{\dot\rho\dot\sigma\dot\tau}
\partial_{\dot\sigma}b_{\lambda\dot\rho}
(\partial_{\dot\lambda}b_{\mu\dot\tau}-\partial_{\dot\tau}b_{\mu\dot\lambda})
\right].
\label{CSt}
\end{eqnarray}
In this so-called ``3+3 formulation'' \cite{Chen:2010jgb} of self-dual gauge theory,
we only need the components $b_{\mu\dm}$ and $b_{\dm\dn}$ of the 2-form potential,
while the components $b_{\mu\nu}$
do not appear until equations of motion are solved.

The Nambu-Poisson bracket $\{\cdot, \cdot, \cdot\}$
is used to define the algebraic structure for gauge symmetry.
In general it satisfies Leibniz rule, fundamental identity,
and here it is defined by
\be
\{f, g, h\} = \eps^{\dm\dn\dlam}\del_{\dm}f \del_{\dn}g \del_{\dlam}h.
\ee
The covariant derivatives are defined by
\bea
{\cal D}_\mu\Phi
&\equiv&\partial_\mu\Phi
-g\{b_{\mu\dot\nu},y^{\dot\nu},\Phi\}
= (\del_{\mu} - gB_{\mu}{}^{\dm}\del_{\dm}) \Phi,
\qquad
(\Phi = X^I, \Psi)
\label{dmu}
\\
{\cal D}_{\dot\mu}\Phi
&\equiv&\frac{g^2}{2}\epsilon_{\dot\mu\dot\nu\dot\rho}
\{X^{\dot\nu},X^{\dot\rho},\Phi\},
\label{ddotmu}
\eea
and the field strengths are defined by
\begin{eqnarray}
{\cal H}_{\lambda\dot\mu\dot\nu}
&=&\epsilon_{\dot\mu\dot\nu\dot\lambda}{\cal D}_\lambda X^{\dot\lambda}
\nonumber\\
&=&H_{\lambda\dot\mu\dot\nu}
-g\epsilon^{\dot\sigma\dot\tau\dot\rho}
(\partial_{\dot\sigma}b_{\lambda\dot\tau})
\partial_{\dot\rho}b_{\dot\mu\dot\nu},\label{h12def}\\
{\cal H}_{\dot1\dot2\dot3}
&=&g^2\{X^{\dot1},X^{\dot2},X^{\dot3}\}-\frac{1}{g}
\nonumber\\
&=&H_{\dot1\dot2\dot3}
+\frac{g}{2}
(\partial_{\dot\mu}b^{\dot\mu}\partial_{\dot\nu}b^{\dot\nu}
-\partial_{\dot\mu}b^{\dot\nu}\partial_{\dot\nu}b^{\dot\mu})
+g^2\{b^{\dot1},b^{\dot2},b^{\dot3}\},
\label{h30def}
\end{eqnarray}
where $H$ is the linear part of the field strength
\begin{eqnarray}
H_{\lambda\dot\mu\dot\nu}
&=&
\partial_{\lambda}b_{\dot\mu\dot\nu}
-\partial_{\dot\mu}b_{\lambda\dot\nu}
+\partial_{\dot\nu}b_{\lambda\dot\mu},\\
H_{\dot\lambda\dot\mu\dot\nu}
&=&
\partial_{\dot\lambda}b_{\dot\mu\dot\nu}
+\partial_{\dot\mu}b_{\dot\nu\dot\lambda}
+\partial_{\dot\nu}b_{\dot\lambda\dot\mu}.
\end{eqnarray}
In the above, we used the notation
\bea
b^{\dm} &\equiv& \frac{1}{2} \eps^{\dm\dn\dlam} b_{\dn\dlam}, \\
X^{\dot\mu}(x) &\equiv&
\frac{x^{\dot\mu}}{g} + b^{\dm}, \\
B_{\mu}{}^{\dm} &\equiv& \eps^{\dm\dn\dlam}\del_{\dn}b_{\mu\dlam}.
\eea
Notice that in the action all appearances of $b_{\mu\dm}$
can be simply expressed in terms of $B_{\mu}{}^{\dm}$ except
the Chern-Simons term $S_{CS}$ (eq.(\ref{CSt})).
While $b^{\dm}$ determines $b_{\dn\dlam}$ uniquely,
$B_{\mu}{}^{\dm}$ does not determine $b_{\mu\dn}$ uniquely.
Nevertheless, with the constraint
\be
\del_{\dm} B_{\mu}{}^{\dm} = 0,
\label{dB0}
\ee
$b_{\mu\dn}$ can be determined by $B_{\mu}{}^{\dm}$
up to a gauge transformation.
Therefore, the physical degrees of freedom represented by
$b_{\dm\dn}$ and $b_{\mu\dn}$
can be equivalently represented by $b^{\dm}$ and $B_{\mu}{}^{\dm}$.

\subsection{Symmetries}

The M5-brane action (eq.(\ref{M5S})) respects
the worldvolume translational symmetry,
the global $SO(2,1)\times SO(3)$ rotation symmetry,
the gauge symmetry for the 2-form gauge potential
and the 6-dimensional ${\cal N}$ = (2, 0) supersymmetry.

\subsubsection{Gauge symmetry}

The Abelian gauge transformation of a 2-form potential $b_{\tilde\mu\tilde\nu}$ is
\be
\d b_{\tilde\mu\tilde\nu} = \del_{\tilde\mu} \Lam_{\tilde\nu} - \del_{\tilde\nu} \Lam_{\tilde\mu},
\ee
where $\Lam_{\tilde\mu}$ is the 1-form gauge transformation parameter.
A remarkable feature of the NP M5-brane theory is that
this higher form gauge symmetry is non-Abelianized.
The gauge transformation laws are
\begin{eqnarray}
\delta_{\Lambda}\Phi
&=&g \kappa^{\dot\mu}\partial_{\dot\mu}\Phi, \label{gt1}
\qquad
(\Phi = X^I, \Psi)
\\
\delta_{\Lambda}b_{\dot\mu\dot\nu}
&=&\partial_{\dot\mu}\Lambda_{\dot\nu}
-\partial_{\dot\nu}\Lambda_{\dot\mu}
+g\kappa^{\dot\lambda}\partial_{\dot\lambda} b_{\dot\mu\dot\nu},
\label{gt2}
\\
\delta_{\Lambda} b_{\mu\dot\mu}
&=&\partial_{\mu}\Lambda_{\dot\mu}
-\partial_{\dot\mu}\Lambda_{\mu}
+g\kappa^{\dot\nu}\partial_{\dot\nu}b_{\mu\dot\mu}
+g(\partial_{\dot\mu}\kappa^{\dot\nu})b_{\mu\dot\nu}, \label{gt4}
\end{eqnarray}
where
\be
\kappa^{\dot\lambda}\equiv
\epsilon^{\dot\lambda\dot\mu\dot\nu}\partial_{\dot\mu}
\Lambda_{\dot\nu}.
\label{def-kappa}
\ee
Eq.(\ref{gt2}) and (\ref{gt4}) can be more concisely expressed in terms of 
$b^{\dm}$ and $B_{\mu}{}^{\dm}$ as
\bea
\d_{\Lambda} b^{\dm} &=& \kappa^{\dm} + g\kappa^{\dn}\del_{\dn} b^{\dm},
\\
\d_{\Lambda} B_{\mu}{}^{\dm} &=&
\del_{\mu}\kappa^{\dm} + g\kappa^{\dn}\del_{\dn}B_{\mu}{}^{\dm}
- g(\del_{\dn}\kappa^{\dm})B_{\mu}{}^{\dn}.
\eea

In terms of $X^I, \Psi, b^{\dm}$ and $B_{\mu}{}^{\dm}$,
the gauge transformation parameter $\Lam_{\mu}$ does not appear,
and all gauge transformations can be expressed solely in terms of $\kappa^{\dm}$,
without referring to $\Lambda_{\dm}$ at all,
as long as one keeps in mind the constraint
\be
\del_{\dm}\kappa^{\dm} = 0.
\ee
As it can be easily seen from eq.(\ref{gt1}),
which is equivalent to a general coordinate transformation
$\d x^{\dm} = \kappa^{\dm}$ by a divergenceless function,
that the manifest gauge symmetry is the volume-preserving diffeomorphism
for the volume-form defined by the large $C$-field background.

\subsubsection{Supersymmetry}

Like $\Psi$,
the SUSY transformation parameter $\eps$ can
be conveniently denoted as an 11D Majorana spinor
satisfying the 6D chirality condition
\be
\Gamma^7 \eps = -\eps.
\ee
The SUSY transformation law is given by
\footnote{
$\eps$ here was denoted as $\eps'$,
and $\d_{\eps}$ here as $\d_{\eps'} + \frac{1}{g}\d_{\chi}$
with $\chi = \G^{\dot{1}\dot{2}\dot{3}}\eps$
in \cite{M52}.
}
\begin{eqnarray}
\delta_\eps X^I
&=&i\ol\epsilon\Gamma^I\Psi,
\label{dX}
\\
\delta_\eps \Psi
&=&{\cal D}_\mu X^I\Gamma^\mu\Gamma^I\epsilon
+{\cal D}_{\dot\mu}X^I\Gamma^{\dot\mu}\Gamma^I\epsilon
-\frac{1}{2}
{\cal H}_{\mu\dot\nu\dot\rho}
\Gamma^\mu\Gamma^{\dot\nu\dot\rho}\epsilon
- {\cal H}_{\dot1\dot2\dot3} \Gamma_{\dot1\dot2\dot3}\epsilon
\nonumber\\&&
-\frac{g^2}{2}\{X_{\dot\mu},X^I,X^J\}
\Gamma^{\dot\mu}\Gamma^{IJ}\epsilon
+\frac{g^2}{6}\{X^I,X^J,X^K\}
\Gamma^{IJK}\Gamma^{\dot1\dot2\dot3}\epsilon,
\label{dPsi}
\\
\delta_\eps b_{\dot\mu\dot\nu}
&=&-i(\ol\epsilon\Gamma_{\dot\mu\dot\nu}\Psi),
\label{db1}
\\
\delta_\eps b_{\mu\dot\nu}
&=&-i\ol\epsilon\Gamma_\mu\Gamma_{\dot\nu}\Psi-ig\ol\epsilon\Gamma_\mu\Gamma_{\dot\lambda}\Psi\del_{\dn}b^{\dlam}
+ig(\ol\epsilon\Gamma_\mu\Gamma_I\Gamma_{\dot1\dot2\dot3}\Psi)
\partial_{\dot\nu}X^I.
\label{db2}
\end{eqnarray}

The above is a linear SUSY transformation.
There is also a nonlinear SUSY
\begin{equation}
\delta_\chi \Psi=\chi,\quad
\delta_\chi X^I=\delta_\chi b_{\dot\mu\dot\nu}
=\delta_\chi b_{\mu\dot\nu}=0.
\label{chitr}
\end{equation}

\subsection{Super Algebra and Central Charges}

The super algebra of the BLG model was discussed in \cite{Lambert,Low}.
The super algebra of the NP M5-brane is essentially the same.
 
Using Noether's theorem, one can calculate the time component of the super current,
\be
\ol{\eps}J^{0}=-(\d_{\eps}\ol{\Psi})\G^{0}\Psi.
\ee
The super charge is the spatial integral of $J^{0}$
\be
Q=\int d^{5}x\: J^{0},
\ee
and it is the generator of supersymmetry transformation, 
so the supersymmetry transformation of fields can be written as
\be
\d_{\eps}\Phi=[\ol{\eps}Q,\Phi].
\ee
One can compute the anticommutator of the super charges as
\be
[\ol{\eps}Q,Q]=\int d^{5}x \; \d_{\eps}J^{0}.
\ee
Finally, the super algebra is of the form
\footnote{This result was first derived in \cite{Low} in the context of BLG model.}
\bea
\{Q,Q\}=2\int d^{5}x\; T_{00}+\sum_{n=0}^{5}\int d^{5}x\;Z_{n},
\eea
where the definition of each term is given in the following.
The first term is the contribution of the energy density
\bea
T_{00}&=&\frac{1}{2}{\cal D}_{0}X^I{\cal D}_{0}X^I
+\frac{1}{2}({\cal D}_{a}X^I)^2
+\frac{1}{4}{\cal H}_{0\dot\mu\dot\nu}{\cal H}_{0}{}^{\dot\mu\dot\nu}
+\frac{1}{4}({\cal H}_{\a\dot\mu\dot\nu})^2\\\nn
&&
+\frac{1}{2}({\cal H}_{\dot1\dot2\dot3})^2
+\frac{g^4}{4}\{X^{\dot\mu}, X^I, X^J\}^2
+\frac{g^4}{12}\{X^I, X^J, X^K\}^2.
\eea
For the sake of convenience to refer to this SUSY algebra later
when we consider BPS states, 
the momentum density ($T_{0a}, a=(\alpha,\dot\mu)$) is included in $Z_n$
together with the central charges.

The symbols $Z_n$ are not defined in accordance with
conventional classification of the central charges based on their tensorial properties.
Instead we classify the terms $Z_n$
according to the number ($n$) of the scalar fields ($X^I$) we choose to turn on 
when we look for BPS states.
More precisely, 
when we turn on $n$ scalar fields,
we can focus on $Z_1, Z_2, \cdots, Z_n$ 
and ignore $Z_{n+1}, Z_{n+2}, \cdots, Z_{5}$.

Let us now describe each term $Z_n$ in order.
The convention of notation here is that
$\alpha, \beta=1, 2$ and $\bar{a}=(0, \dot\mu)$.
We have
\bea
Z_{0}={\cal H}_{0\dot\mu\dot\nu}{\cal H}^{\dot\mu\dot\nu a}
\Gamma^0\Gamma_{a}
-\frac{1}{2}{\cal H}_{\alpha\dot\mu\dot\nu}
{\cal H}_{\beta\dot\lam\dot\rho}
\Gamma^{\dot\mu}\Gamma^{\dot\nu\dot\lam\dot\rho}\Gamma^{\alpha\beta}.
\eea
If we carry out DDR,
the quadratic term in ${\cal H}$ above becomes a quadratic term in 
the D4-brane field strength.
From the D4-brane perspective,
it represents the charge density of D0-branes
on the worldvolume of the D4-brane.
Therefore, from the viewpoint of the M5-brane,
this charge is associated with M-waves propagating on an M5-brane.

Next, we have
\bea
Z_{1}=2{\cal D}_{0}X^{I}{\cal D}_{a}X^{I}
\Gamma^{0}\Gamma^{a}+
\frac{1}{3}{\cal D}_{a}X^I{\cal H}_{bcd}\Gamma^{abcd}\Gamma^I,
\eea
where the self-duality condition
\be
{\cal H}_{\a\b\dm}\equiv-\frac{1}{2}\eps_{0\a\b}\eps_{\dm\dn\dlam}{\cal H}^{0\dn\dlam}
\ee
is used to define ${\cal H}_{\a\b\dm}$.
The $({\cal D}X){\cal H}$ term corresponds to the charge 
associated with the self-dual string.

When two scalar fields are turned on,
we will need to consider $Z_0, Z_1$ and $Z_2$,
where $Z_2$ is defined by
\bea
Z_{2}&=&
{\cal D}_aX^I{\cal D}_bX^J\Gamma^{ab}\Gamma^{IJ}
+2g^2{\cal D}_{\dot\mu}X^I\{X^{\dot\mu}, X^I,X^J\}\Gamma^J\nn\\
&&
+2g^2{\cal D}_0X^I\{X_{\dot\mu},X^I,X^J\}\Gamma^0\Gamma^{\dot\mu}\Gamma^J
+\frac{g^2}{2}{\cal H}_{0\dot\nu\dot\lambda}
\{X_{\dot\mu}, X^I, X^J\}\Gamma^{\dot\mu\dot\nu\dot\lambda}
\Gamma^{IJ}\Gamma^0
\nn\\
&&
-g^2{\cal H}_{\a\dot\mu\dot\nu}\{X^{\dot\nu}, X^I, X^J\}
\Gamma^{\a\dot\mu}\Gamma^{IJ}
-g^2{\cal H}_{\dot1\dot2\dot3}\{X_{\dot\mu}, X^I, X^J\}
\Gamma^{\dot\mu}\Gamma^{IJ}\Gamma_{\dot1\dot2\dot3}.
\eea
The first term ${\cal D}_a X^I{\cal D}_b X^J$ corresponds
to the charge of the 3-brane vortex on the M5-brane worldvolume.

We also have
\bea
Z_{3}&=&
g^2{\cal D}_aX^I\{X_{\dot\mu}, X^J, X^K\}\Gamma^{a\dot\mu}\Gamma^{IJK}
-g^2{\cal D}_{\bar{a}}X^I\{X^I, X^J, X^K\}
\Gamma^{\bar{a}}\Gamma^{JK}\Gamma_{\dot1\dot2\dot3}
\nn\\
&&
-\frac{g^2}{6}{\cal H}_{\a\dot\mu\dot\nu}\{X^I, X^J, X^K\}
\Gamma^{\a\dot\mu\dot\nu}\Gamma^{IJK}\Gamma_{\dot1\dot2\dot3}
\nn\\
&&
+g^4\{X^I, X^J, X_{\dot\mu}\}\{X^I, X^K, X_{\dot\nu}\}
\Gamma^{\dot\mu\dot\nu}\Gamma^{JK}.
\eea
and
\bea
Z_{4}&=&
-\frac{g^2}{3}{\cal D}_{\a}X^I\{X^J, X^K, X^L\}
\Gamma^{\a}\Gamma^{IJKL}\Gamma_{\dot1\dot2\dot3}
\nn\\
&&
+g^4\{X_{\dot\mu}, X^I, X^J\}\{X^I, X^K, X^L\}
\Gamma^{\dot\mu}\Gamma^{JKL}\Gamma_{\dot1\dot2\dot3}
\nn\\
&&
-\frac{g^4}{4}\{X_{\dot\mu}, X^I, X^J\}\{X^{\dot\mu}, X^K, X^L\}\Gamma^{IJKL}.
\eea
The term $\frac{g^2}{3}{\cal D}_{\a}X^I\{X^J, X^K, X^L\}$
is reminiscent of the Basu-Harvey charge,
but with the Nambu bracket replaced by the Nambu-Poisson bracket.

Finally, the last of $Z_n$ is
\bea
Z_{5}=-\frac{g^4}{4}\{X^I, X^J, X^K\}\{X^I, X^L, X^M\}\Gamma^{JKLM}.
\eea
This term is relevant only if we turn on all 5 scalars $X^6, \cdots, X^{10}$.

\section{BPS Solutions}
\label{3}

In this paper, we will only consider {\em bosonic} BPS solutions,
namely those with the fermion field $\Psi = 0$.
The BPS condition is therefore simply that
the SUSY transformation of $\Psi$ eq.(\ref{dPsi}) vanishes
for some SUSY parameters $\eps$.
In this section, we systematically study $1/2$ BPS states
by classifying them according to the number of scalars that are turned on.


A generic feature of the BPS conditions is that
there is a degeneracy of solutions when
the field strength takes a special value
\be
{\cal H}_{\dot{1}\dot{2}\dot{3}} = - \frac{1}{g}.
\ee
This is the value of the field strength ${\cal H}_{\dot{1}\dot{2}\dot{3}}$
when it cancels the background value $1/g$ of the $C$-field.
(Recall that on the M5-brane the $C$-field has to accompanied by
the 3-form field strength $H$ to be gauge-invariant.)
This degeneracy can thus be understood as a result of the breakdown
of the basic assumption that subleading terms of higher order in $1/C$
can be neglected.
We will not consider this degeneracy of solutions in the following.

\subsection{Pure Gauge BPS Configurations}
\label{PureGauge}

Let us start with pure gauge field configurations with 
all scalars turned off, i.e., 
\be
X^I = 0, \qquad \forall I.
\ee
The SUSY transformation law of $\Psi$ is then simplified as
\be
\delta_\eps \Psi =
-\frac{1}{2} {\cal H}_{\mu\dot\nu\dot\lam}
\Gamma^\mu\Gamma^{\dot\nu\dot\lam}\epsilon
- {\cal H}_{\dot1\dot2\dot3} \Gamma_{\dot1\dot2\dot3}\epsilon.
\ee
When the SUSY transformation parameter $\epsilon$
is suitably restricted, 
corresponding BPS states are gauge field configurations
for which the expression above vanishes.
We will focus on $1/2$ BPS states, 
for which $\epsilon$ is restricted to half of its defining space.

Normally, the BPS condition $\d \Psi = 0$
is sufficient to ensure the satisfaction of all equations of motion. 
This is because the preserved SUSY of the BPS states guarantees, 
through the SUSY algebra,
that a certain BPS bound on energy is saturated.
Being the lowest energy states, 
BPS states are stable and expected to satisfy all equations of motion. 
An assumption behind this hand-waving argument
is that the BPS solution under consideration is time-independent.
Thus in principle we need to check that all equations of motion are satisfied
before we claim the discovery of time-dependent BPS states.
In fact, when there are tensor fields, 
we need to check equations of motion when
some time-like components of the tensor fields are turned on.

It turns out that, 
in the theory of NP M5-brane,
there are pure gauge configurations
which preserve half of the supersymmetry
but do not satisfy all equations of motion.
In the explicit examples that we will consider below 
in Sec. \ref{LL2},
this intriguing phenomenon seems to be related with
the particular nature of chiral bosons.

Therefore, in the following we will need to check
the equations of motion for the gauge field \cite{M52}, 
\bea
{\cal D}^{\lam}{\cal H}_{\lam\dm\dn}
+ {\cal D}^{\dlam}{\cal H}_{\dlam\dm\dn} = 0,
\label{EOMnoX1}
\\
{\cal D}^{\lam}\tilde{\cal H}_{\lam\mu\dn}
+ {\cal D}^{\dlam}{\cal H}_{\dlam\mu\dn} = 0,
\label{EOMnoX2}
\eea
when both $X^I$ and $\Psi$ are set to $0$.
Incidentally, the Jacobi identity \cite{M52} is
\be
{\cal D}^{\lam}\tilde{\cal H}_{\lam\mu\nu}
+ {\cal D}^{\dlam}\tilde{\cal H}_{\dlam\mu\nu} = 0.
\ee

As the 6D Lorentz symmetry of the M5-brane is 
broken by the $C$-field background into $SO(2, 1)\times SO(3)$,
there are two types of light-like directions, 
depending on whether the spatial component transforms 
under $SO(2, 1)$ or $SO(3)$.
Without loss of generality, 
we can choose 
\be
x^{\pm} = x^0 \pm x^{\dot{1}}, 
\qquad \mbox{and} \qquad
x^{\pm} = x^0 \pm x^2
\ee
as representatives of the two types of light-like directions.
We will consider BPS light-like solutions for both cases.
They can be interpreted as 
M-waves propagating on the M5-brane worldvolume.
The corresponding solution in the ordinary M5-brane theory 
\cite{ordinary-M5} was found in \cite{Youm}.

\subsubsection{Light-Like BPS Solutions: $x^{\pm} = x^0 \pm x^{\dot{1}}$}
\label{LL1}

For $\eps$ satisfying
\be
\G^{0\dot{1}} \eps = \pm \eps,
\quad \mbox{or equivalently} \quad
\G^{\mp}\eps \equiv (\G^0 \mp \G^{\dot{1}})\eps = 0,
\label{SUSY-LL1}
\ee
the BPS conditions are
\bea
&{\cal H}_{0\dot{2}\dot{3}} = \pm {\cal H}_{\dot{1}\dot{2}\dot{3}},
\qquad
{\cal H}_{1\dot{3}\dot{1}} = \pm {\cal H}_{2\dot{1}\dot{2}},
\qquad
{\cal H}_{2\dot{3}\dot{1}} = \mp {\cal H}_{1\dot{1}\dot{2}},
\label{BPS-LL-1}
\\
&{\cal H}_{1\dot{2}\dot{3}} = {\cal H}_{2\dot{2}\dot{3}}
= {\cal H}_{0\dot{1}\dot{2}} = {\cal H}_{0\dot{3}\dot{1}} = 0.
\label{BPS-LL-2}
\eea
We would like to find configurations $(b_{\dm\dn}, B_{\mu}{}^{\dm})$
that satisfy these conditions.
Their solutions preserve one half SUSY, 
and the associated BPS bound on the energy density is given by
\bea
\left|
{\cal H}_{0\dot2\dot3}{\cal H}_{\dot1\dot2\dot3}
+{\cal H}_{1\dot3\dot1}{\cal H}_{2\dot1\dot2}
-{\cal H}_{2\dot3\dot1}{\cal H}_{1\dot1\dot2}
\right|,
\eea
in agreement with the central charges.

First we impose a gauge fixing condition to reduce
the number of independent components of $b_{\tilde{\mu}\tilde{\nu}}$.
There are 3 components in $b^{\dot{\mu}}$ with 2 independent
gauge transformation parameters
($\Lam_{\dm}$ is equivalent to $\Lam'_{\dm}$ if
their difference is $\del_{\dm}f$ for some function $f$),
so we can impose the gauge fixing condition
\be
b^{\dot{2}} = b^{\dot{3}} = 0,
\label{gauge-fix-b}
\ee
so that 
\be
{\cal H}_{\dot{1}\dot{2}\dot{3}} = \del_{\dot{1}}b^{\dot{1}}.
\ee
In this gauge, 
the field strengths are
\bea
{\cal H}_{\mu\dot{1}\dot{2}} &=& - B_{\mu}{}^{\dot{3}}, \\
{\cal H}_{\mu\dot{3}\dot{1}} &=& - B_{\mu}{}^{\dot{2}}, \\
{\cal H}_{\mu\dot{2}\dot{3}} &=&
\del_{\mu}b^{\dot{1}} - B_{\mu}{}^{\dot{1}}
- g B_{\mu}{}^{\dot{\rho}}\del_{\dot{\rho}}b^{\dot{1}}, \\
{\cal H}_{\dot{1}\dot{2}\dot{3}} &=& \del_{\dot{1}}b^{\dot{1}}.
\eea
The BPS conditions eq.(\ref{BPS-LL-1})--eq.(\ref{BPS-LL-2})
imply that 
we can solve $B_{\mu}{}^{\dot{\mu}}$
in terms of $b^{\dot{1}}$ as
\be
B_0{}^{\dot{1}} =
\frac{(\del_0 \mp \del_{\dot{1}})b^{\dot{1}}}
{1+g{\cal H}_{\dot{1}\dot{2}\dot{3}}}, 
\qquad
B_1{}^{\dot{1}} =
\frac{\del_1 b^{\dot{1}}}{1+g{\cal H}_{\dot{1}\dot{2}\dot{3}}},
\qquad
B_2{}^{\dot{1}} =
\frac{\del_2 b^{\dot{1}}}{1+g{\cal H}_{\dot{1}\dot{2}\dot{3}}},
\label{Bmudot1}
\ee
by assuming that all other components of $B_{\mu}{}^{\dot{\mu}}$ vanish,
that is,
\be
B_{\mu}{}^{\dot{2}} = B_{\mu}{}^{\dot{3}} = 0.
\ee

The consistency condition $\del_{\dm}B_{\mu}{}^{\dm} = 0$
implies that all $B_{\mu}{}^{\dot{1}}$'s are independent of $x^{\dot{1}}$.
This requirement implies that
\bea
\del_{\dot{1}}\left(\frac{\del_a \Phi}{\del_{\dot{1}}\Phi}\right) = 0, 
\eea
where $a = 1, 2, \pm$ and
\be
\Phi \equiv {\cal H}_{\dot{1}\dot{2}\dot{3}} + \frac{1}{g}.
\ee
A class of solutions to this equation is given by
\be
{\cal H}_{\dot{1}\dot{2}\dot{3}} = F(f(x^0, x^1, x^2) + x^{\dot{1}})
\label{LL1-sol}
\ee
for an arbitrary single variable function $F$
and an arbitrary function $f$ depending on $x^{\mu}$ only.
We also checked that (\ref{LL1-sol}) satisfies all equations of motion.
In many supersymmetric field theories there are 
BPS states parametrized by arbitrary functions of light-cone coordinates.
Here we see a much larger class of half BPS solutions 
than we normally expect for a supersymmetry gauge theory.

\subsubsection{Light Like BPS Solutions: $x^{\pm} = x^0 \pm x^2$}
\label{LL2}

Since the rotation symmetry that rotates $x^2$ into $x^{\dot{1}}$
is broken by the $C$-field background,
light-like BPS solutions in the light-like directions
$x^{\pm} \equiv x^0 \pm x^2$
can be different from the light-like BPS solutions introduced above.

For $\eps$ satisfying
\be
\G^{02} \eps = \pm \eps,
\quad
\mbox{or equivalently}
\quad
\G^{\mp} \eps \equiv (\G^0 \mp \G^2) \eps = 0,
\label{G01}
\ee
the BPS conditions are
\be
{\cal H}_{0\dm\dn} = \pm {\cal H}_{2\dm\dn},
\qquad
{\cal H}_{1\dm\dn} = 0,
\qquad
{\cal H}_{\dot{1}\dot{2}\dot{3}} = 0.
\label{Hd1d2d3}
\ee
This solution preserves $1/2$ SUSY. 
Its energy density is bounded by
\bea
\left|\frac{1}{2}{\cal H}_{0\dot\mu\dot\nu}{\cal H}_{2\dot\mu\dot\nu}\right|.
\eea

Since ${\cal H}_{\dot{1}\dot{2}\dot{3}} = 0$, 
we can choose a gauge in which
\bea
b^{\dm}=0.
\eea
From ${\cal H}_{1\dot\mu\dot\nu}$=0, we obtain
\bea
B_{1}{}^{\dot\mu}=0.
\eea
Similarly, 
from ${\cal H}_{0\dot\mu\dot\nu}=\pm{\cal H}_{2\dot\mu\dot\nu}$, 
we get
\bea
B_{0}{}^{\dot\mu}=\pm B_{2}{}^{\dot\mu},
\eea
where $B_0{}^{\dot\mu}$ can be an arbitrary function on M5.
Notice that all configurations obeying the three equations above 
satisfy all BPS conditions and so they preserve half SUSY.
However, as we commented above, 
even when it is a static configuration
(if $B_0{}^{\dm}$ is independent of $x^0$),
we still need to check whether it satisfies all equations of motion.
In fact, a careful examination of the equations of motion reveals 
the fact that only the light-like configurations
\be
B_{0}{}^{\dot\mu}=\pm B_{2}{}^{\dot\mu}= f(x^\pm)
\ee
are genuine BPS states.

\subsection{Single Scalar Field}
\label{SingleScalar}

In addition to the gauge fields,
if a single scalar field, say $X^6$, is turned on, 
the SUSY transformation law eq.(\ref{dPsi}) reduces to
\be
\delta_\eps \Psi
= \left[ {\cal D}_\mu X^6\Gamma^{\mu 6}
+{\cal D}_{\dot\mu}X^6\Gamma^{\dot\mu 6}
-\frac{1}{2} {\cal H}_{\mu\dm\dn} \Gamma^{\mu\dm\dn}
- {\cal H}_{\dot1\dot2\dot3} \Gamma_{\dot1\dot2\dot3}
\right] \epsilon.
\label{dPsi-1}
\ee

Imposing the condition 
\be
\Gamma^{026}\epsilon=\mp\epsilon
\label{G026}
\ee
on the SUSY parameter $\eps$,
we find the BPS conditions
\bea
&{\cal D}_0 X^6=0,
\qquad
{\cal D}_2 X^6=0,
\label{DX0}
\\
&{\cal H}_{0\dot\mu\dot\nu}=0,
\qquad
{\cal H}_{2\dot\mu\dot\nu}=0,
\eea
and
\be
\label{eq:5.1}
{\cal D}_{\hat{\mu}} X^{6} \pm
\frac{1}{6}
\epsilon_{\hat\mu}{}^{\hat\nu \hat\lam \hat\rho}
{\cal H}_{\hat\nu \hat\lam \hat\rho} = 0,
\ee
where $\hat\mu,\hat\nu = 1,\dot{1},\dot{2},\dot{3}$,
and
$\epsilon^{\hat\mu\hat\nu \hat\lam \hat\rho}$
is a totally anti-symmetric tensor with
$\epsilon^{1\dot{1}\dot{2}\dot{3}}=1$.

\subsubsection{Self-Dual String}

A BPS solution of the condition above
is already studied in Ref. \cite{Furuuchi}.
\footnote{
The convention of \cite{Furuuchi} differs from ours
by switching $x_{2}$ with $x_{1}$.
}
This solution preserves one half SUSY
and describes coincident M2-branes ending on an M5-brane
in large $C$-field background.
We expect it to be a deformation of
the self-dual string solution on an M5-brane without $C$-field background 
\cite{soliton-noC}.
The M2-branes extend in the $X^6$ direction
and their intersection with the M5-brane
is the so-called ``self-dual string'' extending along the $x^2$ direction.\\
The energy density is given by
\be
\left|
\frac{1}{6}\epsilon^{\hat\mu\hat\nu\hat\lam \hat\rho}
{\cal D}_{\hat\mu}X^{6}{\cal H}_{\hat\nu \hat\lam \hat\rho}
\right|,
\ee
which can also be expressed as
\be
({\cal D_{\hat\mu}}X^6)^2
\qquad
\mbox{or}
\qquad
\frac{1}{6}({\cal H_{\hat\nu \hat\rho \hat\sigma}})^2.
\ee

We did not find an analytic exact solution to the BPS conditions.
Instead a perturbative solution can be found 
since the zeroth order solution is the same as the self-dual string 
solution on M5-brane without the $C$-field background.
It is
\bea
X^{6}_{(0)} &=& \pm\frac{m}{r^2},\\
{\cal H}_{02\hat{\mu}}^{(0)} &=&
\frac{2m x^{\hat{\mu}}}{r^4},
\label{H02mu}
\eea
where $\hat{\mu} = 1, \dot{1}, \dot{2}, \dot{3}$,
and
\bea
m &\equiv& \frac{k}{(2\pi)^{3/2} (T_{M_{5}})^{1/2}},
\\
r^{2} &\equiv& \sum_{\hat\mu =1}^{\dot{3}} (x_{\hat\mu})^{2}.
\eea
The integer $k$ is a topological charge of the solution
and $T_{M_{5}}$ is the tension of the NP M5-brane \cite{M52}.

While ${\cal H}_{02\mu}$ is given in (\ref{H02mu}),
other components of the field strength ${\cal H}$
are determined by the self-duality relation.
The gauge potential $b$ can be solved 
in the gauge in which
\be
B_{1}{}^{\dot{1}}=B_{1}{}^{\dot{2}}=B_{1}{}^{\dot{3}}=0
\ee
as
\be
b^{\dot\mu}_{(0)} = - \frac{m x^{\dot\mu}}{a^3} A,
\ee
where
\bea
\label{defA}
A &\equiv& (\pm) \frac{\pi}{2}
+ \tan^{-1} \left( \frac{x_1}{a} \right) + \frac{a x_1}{r^2},
\\
a^{2} &\equiv& x^{2}_{\dot{1}} + x^{2}_{\dot{2}} + x^{2}_{\dot{3}}.
\eea
The sign $(\pm)$ in the parenthesis in the definition of $A$ is arbitrary, 
independent of the choice of signs in other expressions.

The next order of the perturbative expansion
\bea
X^6 &=& X^6_{(0)} + gX^6_{(1)} + \cdots, \\
b^{\dm} &=& b^{\dm}_{(0)} + gb^{\dm}_{(1)} + \cdots
\eea
is given by \cite{Furuuchi}
\bea
X^{6}_{(1)}&=&\pm
m^2
\left(
\frac{2x_1}{r^6} + \frac{2A}{ar^4}
\right), \\
b^{\dot\mu}_{(1)}&=& m^{2}x^{\dot\mu}\left\{
\frac{4}{r^6}
-\frac{1}{a^{3}}\left[\frac{\tan^{-1}(\frac{x_{1}}{a})^{2}}{a^{3}}
+\frac{\pi\tan^{-1}(\frac{x_{1}}{a})}{a^{3}}+\frac{a+\pi x_{1}}{r^{4}}
\right.\right.
\nonumber\\
&&+
\left.\left.
\frac{2x_{1}(a^{2}+r^{2})\tan^{-1}(\frac{x_{1}}{a})}{a^{2}r^{4}}
+\frac{a+\pi x_{1}}{a^{2}r^{2}}
+\frac{\pi^2}{4a^3}
\right]
\right\}.
\eea

In Ref. \cite{M52},
the Seiberg-Witten map \cite{Seiberg:1999vs}
was generalized to a map between
the NP M5-brane theory and the ordinary M5-brane theory \cite{ordinary-M5}
with a constant three-form background.
Through the generalized Seiberg-Witten map,
the string soliton solution in the NP M5-brane theory
presented above is shown \cite{Furuuchi} to be
in agreement with that  \cite{soliton-C,Youm}
in the ordinary M5-brane theory 
in constant $C$-field background
up to second order terms in $g$ and $\theta$.

The choice of the $(\pm)$ sign in Eq.(\ref{defA})
corresponds to the choice of the direction of the Dirac string.
At order $g^{0}$, the Dirac string is not physical.
The coupling constant $g$ is a not good expansion parameter
for studying the fate of the Dirac string since $g$ is
associated with the Nambu-Poisson bracket
which has higher derivatives.
Actually the suitable expansion parameter is $gm/a^{3}$.
As a result this expansion is not good when $a^3 \lesssim gm$.
If we want to know more details about the string in the region $a^3 \lesssim gm$,
we need to go beyond the perturbative approximation.
This is still an open problem.

\subsubsection{Self-Dual String in Different Directions}

Instead of preserving the SUSY for $\eps$ satisfying (\ref{G026}), 
one can also consider $\eps$ satisfying another condition
\be
\G^{0\dot{1}6}\eps = \pm \eps.
\ee
This is merely related to the former condition by a rotation, 
but the rotation symmetry is broken by the $C$-field background.
The corresponding self-dual string solution as a BPS state 
can be studied, at least perturbatively. 
The solution is the same as the former at the 0-th order.
It is straightforward, although complicated, to compute higher order terms.
It will be interesting to examine the difference between 
these self-dual strings with different orientations 
to study the effect of the $C$-field background on the self-dual string.
We leave this topic for future study.

\subsubsection{Tilted Brane}

The same BPS condition eqs.(\ref{DX0})--(\ref{eq:5.1})
considered above admits another solution,
in addition to the self-dual string soliton solution.
This solution describes a tilted brane.
We can write down its exact expression as
\bea
{\cal H}_{\dot{1}\dot{2}\dot{3}} &=& h, 
\label{H=h}
\\
X^{6}&=&\pm hx^1, 
\eea
where $h$ is an arbitrary constant.
The gauge potential $b^{\dm}$ can be chosen to be
\be
b^{\dot\mu}=\frac{1}{g}\left((1-h g)^{\frac{1}{3}}-1\right)x^{\dot\mu}
\ee
for this field strength (\ref{H=h}).

\subsection{Double Scalar Fields}

In this section,
we allow two scalar fields to be non-zero.
The BPS condition (the vanishing of eq.(\ref{dPsi})) reduces to
\bea
\left[{\cal D}_\mu X^I\Gamma^{\mu I}
+{\cal D}_{\dot\mu}X^I\Gamma^{\dot\mu I}
-\frac{1}{2} {\cal H}_{\mu\dm\dn} \Gamma^{\mu\dm\dn}
- {\cal H}_{\dot1\dot2\dot3} \Gamma_{\dot1\dot2\dot3}
\right.
\left.
-\frac{g^2}{2}\{X_{\dot\mu}, X^I, X^J\}\Gamma^{\dot\mu}\Gamma^{IJ}
\right]
\epsilon = 0,
\label{dPsi-2}
\eea
when two scalar fields $X^6$, $X^7$ are turned on
in addition to the gauge fields.

\subsubsection{Holomorphic Embedding and M5-M5 Intersection}

Here we study pure scalar field BPS states with two scalar fields turned on.
These include holomorphic embeddings 
of the M5-brane worldvolume in spacetime.
Among them, a special solution represents the configuration of two M5-branes 
intersecting on a 3-dimensional brane.
One of the M5-branes is described as excitations of two scalar fields 
in the worldvolume theory of the other M5-brane.
The solitonic 3-brane solution on M5-brane without $C$-field background 
was discussed in \cite{Howe:1997et}.
Here we study the 3-brane configurations in the large $C$-field background.

Imposing the condition $\Gamma^{671\dot{1}}\epsilon=\pm \epsilon$
on the SUSY parameter $\eps$,
we find the BPS conditions
\bea
0&=&\partial_{1}X^{6}\pm\partial_{\dot{1}}X^{7},
\label{67-1}
\\
0&=&\partial_{\dot{1}}X^{6}\mp\partial_{1}X^{7},
\label{67-2}
\eea
assuming that $X^6, X^7$ depend only on $x^1, x^{\dot{1}}$.
This class of BPS states preserves 1/2 SUSY.
The energy density is bounded by
\bea
\left|\del_{1}X^{6} \del_{\dot{1}}X^{7}
- \del_{\dot{1}}X^{6} \del_{1}X^{7}\right|.
\eea

These BPS states generally describe holomorphic embeddings
of the M5-brane in the 11 dimensional spacetime.
If we use complex coordinates for both target space and base space
\bea
&Z=\pm X^{6}-iX^{7}, \qquad
\bar{Z}=\pm X^{6}+iX^{7},
\\
&z=x^{1}+ix^{\dot{1}}, \qquad
\bar{z}=x^{1}-ix^{\dot{1}},
\eea
BPS conditions eqs.(\ref{67-1}) and (\ref{67-2})
are equivalent to
\bea
\label{eq:hol1}
0&=&\partial_{\bar{z}}Z
,\\
\label{eq:hol2}
0&=&
\partial_{z}\bar{Z}.
\eea
Thus $Z$ is a holomorphic function and $\bar{Z}$
is an anti-holomorphic function.
This holomorphic embedding of M5-brane
has in fact a K$\ddot{a}$hler structure.
The induced metric on the embedding is
\bea
ds^2 =
dx_{1}^{2}+dx_{\dot{1}}^{2}+dX_{6}^{2}+dX_{7}^{2}
= dzd\bar{z}+dZd\bar{Z},
\eea
where we ignored other directions in which
the embedding is trivial.
This is the metric for the K$\ddot{a}$hler potential
\bea
z\bar{z}+Z\bar{Z}.
\eea

Let us give a few examples of this class of solutions.
The first example is given by
\be
\label{eq:sol1}
X^{6}=\pm E_{1}x^{1},
\qquad
X^{7}=-E_{1}x^{\dot{1}}
\ee
where $E_{1}$ is an arbitrary constant.
This solution represents a tilted flat M5-branes.
Due to the presence of the $C$ field background,
this solution is not exactly equivalent to
the vacuum solution through a spacetime rotation.

The second example is given by
\bea
\label{eq:sol2}
X^{6}&=&\pm\frac{E_{2}}{2}\left(x_{1}^{2}-x_{\dot{1}}^{2}\right),\\
\label{eq:sol2-1}
X^{7}&=&-E_{2}x^{1}x^{\dot{1}},
\eea
where $E_{2}$ is an arbitrary constant.
This represents a hyperbolic curve in spacetime.

The last example is given by
\be
Z=\frac{1}{z}, \qquad
\bar{Z} = \frac{1}{\bar{z}}.
\ee
This solution describes the intersection of
two M5-branes at a 3-dimensional space.
The intersecting 3-brane lies along the directions $(0,2,\dot{2},\dot{3})$,
and is located at $x^1 = x^{\dot{1}} = 0$,
where the fields $X^6, X^7$ diverge.
The divergence of $X^6, X^7$
represents another M5-brane extending to infinity in
the directions of $(6, 7)$.

\subsubsection{Deformed Holomorphic Embedding}
\label{deformedholomorphic}

In the above we considered scalar fields with dependence 
on the two directions $x^1$ and $x^{\dot{1}}$.
It is easy to see that analogous results can be easily obtained 
if we replace the two directions by $x^1$ and $x^2$.
But what if we replace them by $x^{\dot{1}}$ and $x^{\dot{2}}$?

Imposing the condition $\Gamma^{67\dot{1}\dot{2}}\epsilon=\pm\epsilon$
on the SUSY transformation parameter $\eps$,
we find the BPS conditions
\bea
\label{eq:t1-0}
{\cal D}_{\mu}X^I &=& 0, \\
\label{eq:t1}
{\cal D}_{\dot{1}}X^{6}\pm{\cal D}_{\dot{2}}X^{7} &=& 0, \\
\label{eq:t1-1}
{\cal D}_{\dot{2}}X^{6}\mp{\cal D}_{\dot{1}}X^{7} &=& 0, \\
\{X^{\dot{1}}, X^6, X^7\} &=& 0, \\
\{X^{\dot{2}}, X^6, X^7\} &=& 0, \\
{\cal H}_{\mu\dm\dn} &=& 0, \\
\label{eq:t1-5}
{\cal H}_{\dot{1}\dot{2}\dot{3}} &=& \pm g^2\{X^{\dot{3}}, X^6, X^7\},\\
{\cal D}_{\dot 3}X^I &=& 0.
\eea

First, let us examine the possibility of describing
two intersecting M5-branes at a 3-brane
located at $x^{\dot{1}}, x^{\dot{2}}$,
without turning on the gauge field.
We obtain from the BPS condition (\ref{eq:t1-5})
\bea
\label{eq:t2}
\partial_{\dot{1}}X^6\partial_{\dot{2}}X^7
=\partial_{\dot{2}}X^6\partial_{\dot{1}}X^7.
\eea
From (\ref{eq:t1}), (\ref{eq:t1-1}) and (\ref{eq:t2}), we find
\bea
-(\partial_{\dot{1}}X^6)^2=(\partial_{\dot{2}}X^6)^2.
\eea
This implies that we must turn on gauge fields
in order to have nontrivial solutions with $g \neq 0$.
In other words, the $C$-field turns on interactions
between the gauge fields and the scalar fields
such that holomorphic embeddings of an M5-brane
are no longer BPS states when 
the complex coordinate $z$ is $x^{\dot{1}} + i x^{\dot{2}}$.

One can find BPS states when the gauge field 
${\cal H}_{\dot{1}\dot{2}\dot{3}}$ is turned on.
The energy density of these BPS states are bounded by
\bea
\left|
{\cal D}_{\dot{1}}X^{6}{\cal D}_{\dot{2}}X^{7}
- {\cal D}_{\dot{2}}X^{6}{\cal D}_{\dot{1}}X^{7}
- g^2{\cal H}_{\dot{1}\dot{2}\dot{3}}\{X^{\dot{3}},X^6,X^7\}
\right|.
\eea
An explicit BPS solution is given by
\bea
X^{6}&=&\pm \left(C_{1}X^{\dot{1}}+C_2 X^{\dot{2}}\right),\\
X^{7}&=&C_{2}X^{\dot{1}}-C_1 X^{\dot{2}},\\
B_{\mu}{}^{\dot{\mu}}&=&0,\\
{\cal H}_{\dot{1}\dot{2}\dot{3}} &=& - \frac{C_1^2 + C_2^2}{g(1 + C_1^2 + C_2^2)},
\label{fconstraint}
\eea
where $C_{1}$ and $C_2$ are the arbitrary constants.
This is again a tilted brane.
But unlike the tilted brane solution above eq.(\ref{eq:sol1}),
the $C$-field background demands that
the tilting of the M5-brane turns on the field strength
${\cal H}_{\dot{1}\dot{2}\dot{3}}$ at the same time.

\section{Conclusion}
\label{4}

In the above we studied BPS states in the NP M5-brane theory,
the low energy effective field theory for an M5-brane in a large $C$-field background.
The large $C$-field background turns on new interactions
on the M5-brane worldvolume through
the Nambu-Poisson structure,
and modifies some of the BPS configurations.
We have only considered 1/2 BPS states, 
and we have not found explicit, nontrivial 1/2 BPS states 
when three or more scalar fields are turned on.
In this section we comment on the significance and implication of our results,
as well as potential future research directions.

In Sec. \ref{PureGauge} we found in passing static configurations 
which do not satisfy all equations of motion but they obey all BPS conditions, 
demonstrating that BPS conditions do not alway imply equations of motion 
even for static states. 
Nevertheless, a hint is hidden in the condition 
of the SUSY transformation parameter:
the condition always involves $\Gamma^0$,
e.g. see (\ref{SUSY-LL1}).
We do not have any counter-example of the widely accepted folklore
when the projection operator defining the preserved SUSY 
does not involve the time component of Dirac $\Gamma$-matrices.
On the other hand, there is no rigorous proof that 
it is impossible to find such counter-examples.

In Sec. \ref{SingleScalar},
we studied self-dual solitons.
The existence of these solutions is already quite interesting by themselves,
as soliton solutions of self-dual gauge field theory 
with {\em non-Abelian} 2-form gauge symmetry.
It would be great if one could find exact self-dual string solutions.
Furthermore, we saw that self-dual solitons on M5-brane depends on 
its orientation due to the $C$-field background. 
It would be interesting to study in more detail this dependence 
and compare it with the effect of the interaction between 
$C$-field and M2-brane worldvolume.

All the known 1/2 BPS states on M5-branes in 
the absence of $C$-field background have their counterparts 
in large $C$-field background, 
although they are sometimes restricted to certain directions.
We expect that some of the BPS state for $C = 0$
will be deformed into 1/4 or less BPS states in large $C$-field background.
For example, we commented in Sec. \ref{deformedholomorphic}
that the holomorphic embedding of M5-brane in spacetime can be
deformed by the $C$-field background if the scalars depend on 
at least two of the $x^{\dm}$ directions.
The only 1/2 BPS states we found in that case as a deformed holomorphic embedding 
was a tilted M5-brane, 
with a field strength ${\cal H}_{\dot{1}\dot{2}\dot{3}}$ that is
consistent with the projection of the $C$-field on the tilted worldvolume.
It will be interesting to study the class of 1/4 BPS states 
which are deformations of holomorphic embeddings due to 
the $C$-field background.
The BPS conditions can be viewed as a natural generalization 
of the notion of holomorphic curves when 
the complex structure is equipped with a 3-form NP structure.

\section*{Acknowledgment}

The authors thank Wei-Ming Chen, Kuo-Wei Huang, Yen-Ta Huang, Fech Scen Khoo,
Yi-Chuan Lu and Chen-Pin Yeh
for helpful discussions. 
This work is supported in part by
the National Science Council.

\end{document}